\documentclass[12pt,letterpaper]{article}
\usepackage{t1enc}
\usepackage[latin1]{inputenc}
\usepackage[english]{babel}
\usepackage{graphicx}
\usepackage{amsmath}
\usepackage{amsfonts}

\renewcommand{\baselinestretch}{1.75}

\begin{document}

\title{Multivariate Nonnegative Trigonometric Sums Distributions for High-Dimensional Multivariate Circular Data}
\renewcommand{\baselinestretch}{1.00}
\author{Fern\'andez-Dur\'an, Juan Jos\'e$$ and Mar\'ia Mercedes Gregorio-Dom\'inguez$^{*}$ \\
ITAM\\
CDMX, M\'exico \\
$^*$ Corresponding Author: \\
E-mail: \texttt{mercedes@itam.mx}}
\date{}
\maketitle

\renewcommand{\baselinestretch}{1}

\begin{abstract}
Fern\'andez-Dur\'an and Gregorio-Dom\'inguez (2014) defined a family of probability distributions for a vector of circular random variables by considering multiple nonnegative trigonometric sums. These distributions are highly flexible and can present numerous modes and skewness. Several operations on these multivariate distributions were translated into operations on the vector of
parameters; for instance, marginalization involves calculating the eigenvectors and eigenvalues of a matrix, and
independence among subsets of the vector of circular variables translates to a Kronecker product of the corresponding subsets of the vector of parameters.
Furthermore, it was demonstrated that the family of multivariate circular distributions based on nonnegative trigonometric sums is closed under marginalization and conditioning, that is, the marginal and conditional densities of any order are also members of the family. The derivation of marginal and conditional densities from the joint multivariate density is important when applying this model in practice to real datasets. A goodness-of-fit test based on the characteristic function and an alternative parameter estimation algorithm for high-dimensional circular data was presented and applied to a real dataset on the daily times of occurrence of maxima and minima of prices in financial markets.
\end{abstract}


\newpage

\renewcommand{\baselinestretch}{1.75}

\section{Introduction}

This study considers the properties of the multivariate nonnegative trigonometric sums (MNNTS) distributions developed by
Fern\'andez-Dur\'an and Gregorio-Dom\'inguez (2014), such as conditional and marginal distributions and the conditions for independence among subsets of the vector of circular random
variables. A circular (angular) random variable is defined as one where the support
of its probability density function is the unit circle, and it must be a function of period 2$\pi$, that is, if $f(\theta)$ represents the density function
of a circular random variable, $\theta$, then, $f(\theta + 2k\pi)=f(\theta)$, where
$k$ is an integer. A multivariate circular random vector is a vector
in which each component is a circular random variable, $\underline{\theta}=(\theta_1, \theta_2, \ldots, \theta_d)^\top$, where $\theta_1, \theta_2, \ldots, \theta_d$ are circular random variables.

Multivariate circular data are available in numerous disciplines. The dihedral angles in a protein, wind directions recorded at different monitoring stations, time of occurrence of different diseases, and time of flowering of different plant species are some instances. In numerous applications, multivariate circular data comprise the time of occurrence of different events. The support of the domain of the distribution of a multivariate circular random vector is a hypertorus. The bivariate von Mises model was developed by Mardia (1975a; see also Mardia and Jupp 2000) that was extended later to a $d$-dimensional torus by Mardia et al. (2008) but all these models suffer from an unknown normalizing constant or a normalizing constant in the form of an infinite sum and their marginal densities lack a closed expression and in many cases are restricted to be symmetric. Johnson and Wehrly (1977) and Wehrly and Johnson (1980) presented bivariate circular models based on a decomposition of the bivariate cumulative distribution that is equivalent to a copula (Nelsen 1999) used by Fern\'andez-Dur\'an (2007) to construct bivariate circular and circular-linear probability models based on univariate nonnegative trigonometric sums (Fern\'andez-Dur\'an 2004; Fern\'andez-Dur\'an and Gregorio-Dom\'inguez 2012). Other bivariate circular models
are presented by Singh et al. (2002), Lennox et al. (2009), Mardia et al. (2007),
and Shieh and Johnson (2005). In the multivariate circular case, Kim et al. (2016) extended the bivariate copula decomposition of Johnson and Wehrly (1977) and Wehrly and Johnson (1980) to the multivariate case. The d-dimensional density function of the model proposed by Kim et al. (2016) is defined as follows:
\begin{equation}
f(\underline{\theta})=(2\pi)^p \prod_{j=1}^p \left\{ g_j \left(  2\pi \sum_{k=j}^d F_k(\theta_k) \right) \right\} \prod_{k=1}^d f_k(\theta_k)
\end{equation}
where $1 \le p \le d-1$. The functions $f_k$ and $g_j$ are circular density functions and the functions $F_k$ are the corresponding circular distribution functions. The circular random variables $\theta_{p+1}, \ldots, \theta_{d}$ are mutually independent. The total number of parameters is equal to the sum of the number of parameters of the $g$ and $f$ functions. For instance, when considering two-parameter von Mises distributions for all the $g$ and $f$ functions, there are a total of 2$p$+2$d$ free parameters. For a particular application, selecting marginal densities $f$ could involve circular densities with a larger number of parameters. In Kim et al. (2016), their model was compared to the MNNTS model, and the latter proved to be the best Akaike Information Criterion (AIC) model when fitted to a dataset of three dihedral angles in a protein. Nadehi et al. (2021) proposed using multivariate wrapped models to fit data on a $d$-dimensional torus. Mardia et al. (2012)
considered mixtures of multivariate circular distributions that are extensions
of the univariate von Mises model with a total of $q\left(\frac{d^2+3d}{2}\right)$ free parameters where $q$ is the number of components in the mixture and $d$ is the dimension of the circular vector. For instance, for a 6-dimensional circular random vector with eight components in the mixture, there is a total of 216 parameters. Another model that is an extension of the univariate von Mises model is presented by Mardia and Voss (2014). Using mixtures of multivariate wrapped models that can allow for an unspecified large number of modes is still not developed. In previous multivariate circular models, the authors did not present goodness-of-fit tests for their models. Fern\'andez-Dur\'an and Gregorio-Dom\'inguez (2014) extended the univariate nonnegative trigonometric model to the multivariate case in which the parameters
are estimated by maximum likelihood (refer to Fern\'andez-Dur\'an and Gregorio-Dom\'inguez 2010) and computationally implemented using the R software
(R Development Core Team 2020) in the CircNNTSR library (Fern\'andez-Dur\'an and
Gregorio-Dom\'inguez 2012, 2016). We refer to this family of multivariate
circular distributions as the MNNTS. The MNNTS models enable the fitting of multivariate circular datasets that present multimodality and/or skewness. A modified Newton optimization algorithm on manifolds efficiently obtains maximum likelihood estimates of the parameters of the MNNTS models (see Fern\'andez-Dur\'an and Gregorio-Dom\'inguez 2010, 2016). Thus, the maximum likelihood estimation can be used to develop tests of conditional independence and uniformity. The multivariate circular uniform distribution is a special case of the MNNTS models when componentwise $\underline{M}=\underline{0}$. The family of MNNTS models is nested, that is, if componentwise $\underline{M} \le \underline{M}^*$, then the MNNTS model with $\underline{M}$ is a particular case of the MNNTS model with $\underline{M}^*$. The density of MNNTS models is fully known and does not have an unknown normalizing constant, unlike other multivariate circular models. This study demonstrates that the MNNTS family is closed under marginalization and conditioning, that is, the marginal and conditional densities of any order are also MNNTS densities. The main drawback of MNNTS models, as with many mixtures of multivariate circular distributions models, is their many parameters, necessitating a large amount of data for fitting. Furthermore, if the multivariate circular observations are highly concentrated in few regions then, an MNNTS model with a large $\underline{M}$ will be required. Although, highly concentrated multivariate circular observations can also be modeled by classical linear models because the periodicity constraint of the multivariate circular distributions could be ignored. The MNNTS probability density function for multivariate circular random vectors is defined as follows:
\begin{eqnarray}
f_{12 \cdots d}(\underline{\theta})=||\underline{c}^H\underline{e}||^2 & = & \underline{c}^H\underline{e}\underline{e}^H\underline{c} \\\nonumber
& = & \sum_{k_1=0}^{M_1}\sum_{k_2=0}^{M_2} \cdots \sum_{k_d=0}^{M_d} \sum_{m_1=0}^{M_1}\sum_{m_2=0}^{M_2} \cdots \sum_{m_d=0}^{M_d} c_{k_1k_2 \cdots k_d}\bar{c}_{m_1m_2 \cdots m_d}e^{\sum_{s=1}^{d}i(k_s-m_s)\theta_s},  \\\nonumber
\label{mnntssum}
\end{eqnarray}
where $i=\sqrt{-1}$, $c=c_R + ic_I$, $c_R$ and $c_I$ are the real and imaginary parts of complex number $c$, and $\bar{c}=c_R-ic_I$ is the conjugate of the complex number $c$. The complex vector, $\underline{c}$, is the vector of parameters. The complex vector, $\underline{e}$, contains the multivariate trigonometric moments defined as $e^{\sum_{s=1}^{d}r_s\theta_s}$ for integer values $r_1, r_2, \ldots, r_d$. Both vectors have the dimension of $\prod_{s=1}^{d}(M_s+1)$.
The vector of parameters, $\underline{c}$, must satisfy the following constraint:
\begin{equation}
||\underline{c}||^2 = \sum_{k_1=0}^{M_1}\sum_{k_2=0}^{M_2} \cdots \sum_{k_d=0}^{M_d} ||c_{k_1 k_2 \cdots k_d}||^2 = \frac{1}{(2\pi)^d}.
\end{equation}
Given this constraint, the first element of the $\underline{c}$ vector, $c_{00 \cdots 0}$, is a nonnegative real number, and the parameter
space is a complex hypersphere with a dimension of $\prod_{s=1}^{d}(M_s+1)$, which is isomorphic to a real hypersphere with a dimension of $2\prod_{s=1}^{d}(M_s+1)-1$ by taking the real and imaginary parts of the complex numbers in $\underline{c}$. Additionally, for the identifiability of the parameters, $c_{0 \ldots 0}$ should be greater than or equal to $||c_{M_1M_2 \ldots M_d}||$ because the model with the conjugate of $\underline{c}$ written in reverse order outputs the same model as $\underline{c}$.

Given the definition of the multivariate density function in Equation \ref{mnntssum}, it is easy to obtain the expression for the marginal distributions; however, the parameters of the marginal distributions for any dimension remain unclear. In addition, the conditional densities of a subset of
the circular variables, given other subsets of the joint vector, are unclear at the first instance.

The remainder of this paper is organized as follows. Section 2 presents alternative forms of writing the density function in Equation \ref{mnntssum} that are better suitted for deriving the parameters of marginal and conditional distributions. Section 3 describes the development of marginal distributions. Section 4 provides the conditions for independence among the elements
of the multivariate circular vector, and Section 5 details the derivation of the
conditional distributions. The derivation of the marginal and conditional distributions is developed for the joint bivariate cases, although their generalization to any dimension is direct. In Section 6, a goodness-of-fit test for MNNTS models is developed by considering the characteristic function. In Section 7, we propose an efficient algorithm to estimate high-dimensional circular data parameters, an alternative to maximum likelihood estimation, which can be slow or fail to converge in such cases. Section 8 presents the application of the results described in the
previous sections to a real dataset on the daily times of occurrence of maxima and minima of prices in financial markets. Finally, Section 9 concludes this study.

\section{MNNTS Distributions}

Consider a vector of circular random variables, $\underline{\theta}=(\theta_1, \theta_2, \ldots, \theta_d)^{\top}$, which is distributed as an MNNTS distribution (Fern\'andez-Dur\'an 2007; Fern\'andez-Dur\'an and Gregorio-Dom\'inguez 2014). Subsequently, to perform efficient (numerical) calculations, the
density function of $\underline{\theta}$ in Equation \ref{mnntssum} can be written in terms of Kronecker products as follows:
\begin{equation}
f_{12 \cdots d}(\underline{\theta})= \underline{c}^H\underline{e}\underline{e}^H\underline{c} =
\underline{c}^H \left(\bigotimes_{s=1}^{d}\underline{e}_{s}\bigotimes_{s=1}^{d}\underline{e}_{s}^{H}\right)\underline{c} =
\underline{c}^H \left(\bigotimes_{s=1}^{d}\underline{e}_{s}\underline{e}_{s}^{H}\right)\underline{c},
\label{mnntskronecker}
\end{equation}
where $\underline{e}_s=(1,e^{i\theta_s},e^{2i\theta_s}, \ldots, e^{M_{s}i\theta_s})^\top$ is the vector of the trigonometric moments of the s-th circular random variable in the random vector, $\underline{\theta}$. This result is based on the successive applications of the following property of Kronecker products:
\begin{equation}
(A \otimes B)(C \otimes D) = (AC) \otimes (BD),
\label{kroneckerprop}
\end{equation}
where $A$, $B$, $C$, and $D$ are general matrices for which the products are defined.
This expression can also be used in computational programs to perform numerical calculations more efficiently. Furthermore, the set of indices of the vector of parameters, $\underline{c}$, is obtained using the Kronecker product,
$(0,1, \ldots, M_1) \bigotimes (0,1, \ldots, M_2) \bigotimes \cdots \bigotimes (0,1, \ldots, M_d)$, for a $d$-dimensional MNNTS model. The order of the elements
of the circular vector $\underline{\theta}$ can be modified to simplify the derivations presented in the following sections. When expressed in terms of cosine and sine functions, the MNNTS density function is a weighted sum of cosines and sines of integer linear combinations of the elements of the circular vector. For instance, the bivariate MNNTS model with $\underline{M}=(1,1)$ and $\underline{c}=(c_{00},c_{01},c_{10},c_{11})^\top$ has the following expression of its density function in terms of cosine and sine terms:
\begin{eqnarray*}
f_{12}(\theta_1,\theta_2) & = & \frac{1}{(2\pi)^2} + 2Re(c_{00}\bar{c}_{11})\cos(\theta_1+\theta_2) + \\
 & & 2Im(c_{00}\bar{c}_{11})\sin(\theta_1+\theta_2) + 2Re(c_{00}c_{10}+c_{11}\bar{c}_{01})\cos(\theta_1) + \\
 & & 2Im(c_{00}c_{10}+c_{11}\bar{c}_{01})\sin(\theta_1) + 2Re(c_{00}c_{01}+c_{11}\bar{c}_{10})\cos(\theta_2) + \\
 & & 2Im(c_{00}c_{01}+c_{11}\bar{c}_{10})\sin(\theta_2) + 2Re(c_{10}\bar{c}_{01})\cos(\theta_1-\theta_2) + \\
 & & 2Im(c_{10}\bar{c}_{01})\sin(\theta_1-\theta_2)
\end{eqnarray*}
where $Re(a)$ and $Im(a)$ are the real and imaginary parts of the complex number $a$, respectively.

\section{Marginal Distributions}

The expression for the marginal distribution of any dimension of an MNNTS distribution is obtained by integrating the marginalizing components. For the bivariate joint MNNTS distribution in terms of the sum,
\begin{eqnarray}
\nonumber f_1(\theta_1)=\int_{0}^{2\pi}f_{12}(\theta_1,\theta_2)d\theta_2 & = &
\int_{0}^{2\pi} \sum_{k_1=0}^{M_1}\sum_{k_2=0}^{M_2}\sum_{m_1=0}^{M_1}\sum_{m_2=0}^{M_2} c_{k_1k_2}^{(12)}\bar{c}_{m_1m_2}^{(12)}e^{i(k_1-m_1)\theta_1 + i(k_2-m_2)\theta_2}d\theta_2 \\\nonumber
& = & \sum_{k_1=0}^{M_1}\sum_{k_2=0}^{M_2}\sum_{m_1=0}^{M_1}\sum_{m_2=0}^{M_2} c_{k_1k_2}^{(12)}\bar{c}_{m_1m_2}^{(12)}e^{i(k_1-m_1)\theta_1}\int_{0}^{2\pi} e^{i(k_2-m_2)\theta_2}d\theta_2 \\
& = & 2\pi \sum_{k_1=0}^{M_1}\sum_{m_1=0}^{M_1}\left(\sum_{m_2=0}^{M_2} c_{k_1m_2}^{(12)}\bar{c}_{m_1m_2}^{(12)}\right)e^{i(k_1-m_1)\theta_1}.
\end{eqnarray}
Subsequently, if the general sum form of the marginal distribution of $\theta_1$ is
\begin{equation}
f_1(\theta_1)=\sum_{k_1=1}^{M_1}\sum_{m_1=1}^{M_1}c_{k_1}^{(1)}\bar{c}_{m_1}^{(1)}e^{i(k_1 - m_1)\theta_1},
\end{equation}
we can obtain
\begin{equation}
c_{k_1}^{(1)}\bar{c}_{m_1}^{(1)} = 2\pi\sum_{m_2=0}^{M_2} c_{k_1m_2}^{(12)}\bar{c}_{m_1m_2}^{(12)},
\end{equation}
where $\underline{c}^{(1)}$ is the vector of the parameters of marginal density $f_1$. The expression for $f_1(\theta_1)$ is easy to obtain; however, obtaining the expression of the vector of the parameters of the univariate distribution, $\underline{c}^{(1)}=(c_0^{(1)}, c_1^{(1)}, \ldots, c_{M_1}^{(1)})^\top$, in terms of the bivariate distribution vector of parameters, $\underline{c}^{(12)}=(c_{00}^{(12)},c_{01}^{(12)}, \ldots, c_{0M_2}^{(12)}, \ldots, c_{M_10}^{(12)}, \ldots, c_{M_1M_2}^{(12)})^\top$, is a difficult task that involves matrix algebra. Considering the joint bivariate MNNTS distribution for vector $\underline{\theta}=(\theta_1,\theta_2)^\top$, we obtain the marginal distribution of $\theta_1$.
\begin{eqnarray}
\nonumber f_1(\theta_1)=\int_{0}^{2\pi}f_{12}(\theta_1,\theta_2)d\theta_2 &=&
\int_{0}^{2\pi}\underline{c}^{(12)H}\left((\underline{e}_1\underline{e}_{1}^{H})\otimes (\underline{e}_2\underline{e}_{2}^{H})\right)\underline{c}^{(12)}d\theta_2 \\
&=& \underline{c}^{(12)H}\left((\underline{e}_1\underline{e}_{1}^{H})\otimes \left(\int_{0}^{2\pi}\underline{e}_2\underline{e}_{2}^{H}d\theta_2\right)\right)\underline{c}^{(12)}
\end{eqnarray}
Thus,
\begin{equation}
f_1(\theta_1)=
2\pi\underline{c}^{(12)H}\left((\underline{e}_1\underline{e}_{1}^{H})\otimes \mathbb{I}_{M_2+1} \right)\underline{c}^{(12)}=
2\pi\underline{c}^{(12)H}\left((\underline{e}_1 \otimes \mathbb{I}_{M_2+1}\right)\left(\underline{e}_{1}^{H}\otimes \mathbb{I}_{M_2+1} \right)\underline{c}^{(12)}.
\end{equation}
If the bivariate MNNTS vector parameter is written as
\begin{equation}
\underline{c}^{(12)}=(\underline{c}_{\bullet 0}^{(12)},\underline{c}_{\bullet 1}^{(12)}, \ldots, \underline{c}_{\bullet M_2}^{(12)})^\top
\end{equation}
with $\underline{c}_{\bullet m_2}^{(12)}=(c_{0m_2}^{(12)},c_{1m_2}^{(12)}, \ldots, c_{M_1m_2}^{(12)})^\top$ for $m_2=0, 1, \ldots, M_2$, then
\begin{eqnarray}
\nonumber f_1(\theta_1) & = &
2\pi\underline{c}^{(12)H}\left((\underline{e}_1 \otimes \mathbb{I}_{M_2+1}\right)\left(\underline{e}_{1}^{H}\otimes \mathbb{I}_{M_2+1} \right)\underline{c}^{(12)} \\
\nonumber & = & 2\pi\sum_{m_2=0}^{M_2}{c}_{\bullet m_2}^{(12)H}\underline{e}_1\underline{e}_{1}^{H}{c}_{\bullet m_2}^{(12)} \\
 & = & 2\pi \underline{e}_{1}^{H} \left( \sum_{m_2=0}^{M_2}{c}_{\bullet m_2}^{(12)}{c}_{\bullet m_2}^{(12)H} \right) \underline{e}_{1}.
\end{eqnarray}
The spectral decomposition of $C_{\bullet 2} = 2\pi \sum_{m_2=0}^{M_2}{\underline{c}}_{\bullet m_2}^{(12)}{\underline{c}}_{\bullet m_2}^{(12)H}$ satisfies
\begin{equation}
C_{\bullet 2} = \sum_{m_2=0}^{M_2} p_{m_2}{\underline{c}}_{\bullet m_2}^{*(12)}{\underline{c}}_{\bullet m_2}^{*(12)H}.
\label{eqmarginal2b}
\end{equation}
Subsequently, the marginal distribution of $\theta_1$ is a mixture of the univariate NNTS densities:
\begin{equation}
f_1(\theta_1) = \underline{e}_{1}^{H} \left( \sum_{m_2=0}^{M_2}p_{m_2}{\underline{c}}_{\bullet m_2}^{*(12)}{\underline{c}}_{\bullet m_2}^{*(12)H} \right) \underline{e}_{1} =
\sum_{m_2=0}^{M_2}p_{m_2} {\underline{c}}_{\bullet m_2}^{*(12)H}\underline{e}_{1}\underline{e}_{1}^{H}{\underline{c}}_{\bullet m_2}^{(12)*}
\label{marginaleq1}
\end{equation}
where the mixture probabilities, $p_0$, $p_1$, $\ldots$, $p_{M_2}$, and mixture parameter vectors, $\underline{c}_{\bullet 0}^{*(12)}$, $\underline{c}_{\bullet 1}^{*(12)}$ $\ldots$,
$\underline{c}_{\bullet M_2}^{*}$, corresponding to the eigenvalues and eigenvectors of matrix $C_{\bullet 2}$. This result generalizes to the marginals of any dimension of an MNNTS distribution, that is, if $\underline{\theta}=(\underline{\theta}_R,
\underline{\theta}_{R^c})^\top$, where $R$ is the set of indices of the circular random variables for which we need its marginal MNNTS distribution and $R^c$ is the complement set of $R$, then the marginal distribution of $\underline{\theta}_R$ is a mixture of MNNTS distributions.

\section{Independence}

In the simplest case of a bivariate MNNTS distribution for $\underline{\theta}=(\theta_1,\theta_2)^\top$, the circular random variables, $\theta_1$ and $\theta_2$, are independent if and only if the joint vector of parameters, $\underline{c}^{(12)}$, is the Kronecker product of the marginal vector of parameters
$\underline{c}^{(1)}$ and $\underline{c}^{(2)}$, that is, $\underline{c}^{(12)}=\underline{c}^{(1)} \otimes \underline{c}^{(2)}$. The proof uses the result of Equation \ref{kroneckerprop}. If $\underline{c}^{(12)}=\underline{c}^{(1)} \otimes \underline{c}^{(2)}$,
\begin{equation}
f_{12}(\theta_1,\theta_2)=
\underline{c}^{(12)H}\left((\underline{e}_1\underline{e}_{1}^{H}) \otimes (\underline{e}_2\underline{e}_{2}^{H})\right)\underline{c}^{(12)}=
(\underline{c}^{(1)} \otimes \underline{c}^{(2)})^H\left((\underline{e}_1\underline{e}_{1}^{H}) \otimes (\underline{e}_2\underline{e}_{2}^{H})\right)(\underline{c}^{(1)} \otimes \underline{c}^{(2)}).
\label{independenceeq1}
\end{equation}

Repeated applications of the property of Kronecker products in Equation \ref{kroneckerprop} obtain
\begin{equation}
f_{12}(\theta_1,\theta_2)=\left[(\underline{c}^{(1)H}(\underline{e}_1\underline{e}_{1}^{H})) \otimes (\underline{c}^{(2)H}(\underline{e}_2\underline{e}_{2}^{H})) \right](\underline{c}^{(1)} \otimes \underline{c}^{(2)})=
\left[\underline{c}^{(1)H}(\underline{e}_1\underline{e}_{1}^{H})\underline{c}^{(1)} \right] \otimes \left[\underline{c}^{(2)H}(\underline{e}_2\underline{e}_{2}^{H})\underline{c}^{(2)} \right],
\label{independenceeq2}
\end{equation}
which is equal to $f_1(\theta_1)f_2(\theta_2)$ because the Kronecker product of the two scalars is equal to a simple product. The proof that if $\theta_1$ and $\theta_2$ are independent, then the joint parameter vector is the product of the marginal parameter vectors obtained from the previous proof starting from the end.
This result can be generalized to MNNTS of any dimension. Let $\underline{\theta}=(\underline{\theta}_R,\underline{\theta}_{R^c})^\top$, then $\underline{\theta}_R$ is independent of $\underline{\theta}_{R^c}$ if and only if $\underline{c}^{(R \bigcup R^c)}=\underline{c}^{(R)} \otimes \underline{c}^{(R^c)}$. Using this result, we can construct
a likelihood ratio test for independence to determine the independence among relevant subsets of the vector of circular random variables, $\underline{\theta}$.
In terms of the decomposition of the univariate marginal distribution as a mixture of distributions in Equation \ref{marginaleq1}, $\underline{\theta}_R$ is independent of $\underline{\theta}_{R^c}$
when there is only one element in the mixture, that is, the eigenvalues (mixing probabilities) are equal to zero except for the first, which is equal to one.

Based on the definition of the MNNTS density, it is clear that the elements of the vector of circular random variables, $\underline{\theta}=(\theta_1,\theta_2, \ldots, \theta_d)^\top$, are exchangeable if they are equally distributed and then all the components of the vector of the number of terms in the sum, $\underline{M}=(M_1, M_2, \ldots, M_d)$, are equal, that is,
$M_k=M$ for $k=1, 2, \ldots, d$.

\section{Conditional Distributions}

The conditional distributions of an MNNTS distribution are also MNNTS distributions, that is, the MNNTS family of distributions is closed under conditioning.
For the bivariate case, the conditional distribution of $\theta_1$, given $\theta_2=\theta_{2}^{*}$, is obtained as follows:
\begin{equation}
f_{1 \mid 2}(\theta_1 \mid \theta_2=\theta_{2}^{*})=\frac{f_{12}(\theta_{1},\theta_2^{*})}{f_2(\theta_{2}^{*})}.
\end{equation}
If $f_{12}(\theta_{1},\theta_2^{*})$ for a fixed value of $\theta_2=\theta_{2}^{*}$,
\begin{equation}
f_{12}(\theta_{1},\theta_2^{*})=\underline{c}^{(12)H}\left[(\underline{e}_{1}\underline{e}_{1}^{H}) \otimes (\underline{e}_{2}^{*}\underline{e}_{2}^{*H}) \right]\underline{c}^{(12)}=
\underline{c}^{(12)H}\left[(\underline{e}_{1}\underline{e}_{1}^{H}) \otimes (\underline{e}_{2}^{*}\underline{e}_{2}^{*H}) \right]\underline{c}^{(12)}.
\end{equation}
Because for an $m \times n$ $A$ and a $p \times q$ $B$ matrices, $A \otimes B =(A \otimes \mathbb{I}_p)(\mathbb{I}_n \otimes B) = (\mathbb{I}_m \otimes B)(A \otimes \mathbb{I}_q)$ where $\mathbb{I}_s$ denotes the $s$-by-$s$ identity matrix,
\begin{eqnarray}
\nonumber f_{12}(\theta_{1},\theta_2^{*}) & = &
\underline{c}^{(12)H}\left[\left[(\underline{e}_{1}\underline{e}_{1}^{H})\otimes \mathbb{I}_{M_2 + 1}\right]
\left[\mathbb{I}_{M_1+1} \otimes (\underline{e}_{2}^{*}\underline{e}_{2}^{*H})\right] \right]\underline{c}^{(12)} \\
\nonumber & = & \underline{c}^{(12)H}\left[\left[(\underline{e}_{1}\underline{e}_{1}^{H})\otimes (\mathbb{I}_{M_2 + 1}\mathbb{I}_{M_2 + 1})\right]
\left[(\mathbb{I}_{M_1+1}\mathbb{I}_{M_1+1}) \otimes (\underline{e}_{2}^{*}\underline{e}_{2}^{*H})\right] \right]\underline{c}^{(12)} \\
\nonumber & = & \underline{c}^{(12)H}\left[(\underline{e}_{1} \otimes \mathbb{I}_{M_2 + 1})(\underline{e}_{1}^{H} \otimes \mathbb{I}_{M_2 + 1})
(\mathbb{I}_{M_1+1} \otimes \underline{e}_{2}^{*})(\mathbb{I}_{M_1+1} \otimes \underline{e}_{2}^{*H}) \right]\underline{c}^{(12)} \\
\nonumber & = & \underline{c}^{(12)H}\left[(\underline{e}_{1} \otimes \mathbb{I}_{M_2 + 1})
(\underline{e}_{2}^{*} \otimes 1)
(1 \otimes \underline{e}_{1}^{H})
(\mathbb{I}_{M_1+1} \otimes \underline{e}_{2}^{*H}) \right]\underline{c}^{(12)} \\
\nonumber & = &  \underline{c}^{(12)H}\left[
(\mathbb{I}_{M_1 + 1} \otimes \underline{e}_{2}^{*})
(1 \otimes \underline{e}_{1} )
(1 \otimes \underline{e}_{1}^{H})
(\mathbb{I}_{M_1+1} \otimes \underline{e}_{2}^{*H}) \right]\underline{c}^{(12)} \\
& = & \underline{c}^{*(12)H}(\underline{e}_{1}\underline{e}_{1}^{H})\underline{c}^{*(12)},
\end{eqnarray}
where $\underline{c}^{*(12)}=(\mathbb{I}_{M_1+1} \otimes \underline{e}_{2}^{*H})\underline{c}^{(12)}$. Subsequently, the vector of parameters of the conditional distribution,
$\underline{c}^{(1 \mid 2)}$, satisfies
\begin{equation}
\underline{c}^{(1 \mid 2)} = \frac{\underline{c}^{(12)*}}{\sqrt{f_2(\theta_{2}^{*})}} =
\frac{(\mathbb{I}_{M_1+1} \otimes \underline{e}_{2}^{*H})\underline{c}^{(12)}}{\sqrt{f_2(\theta_{2}^{*})}}
\end{equation}
because $\sqrt{f_2(\theta_{2}^{*})}$ is the constant of proportionality. Geometrically, the conditional parameter vector, $\underline{c}^{(1 \mid 2)}$, is obtained by rotating the joint parameter vector by $\theta_2^*$ through $\underline{e}_{2}^{*H}$ and then normalizing it. This result can be generalized to the multivariate case by considering $\underline{\theta}=(\underline{\theta}_{C^c},\underline{\theta}_{C})^\top$, where $C$ is the set of indexes of the conditioning circular variables, given
the known values, $\underline{\theta}_{C}=\underline{\theta}_{C}^{*}$. The conditional distribution of $\underline{\theta}_{C^c}$, given $\underline{\theta}_{C}=\underline{\theta}_{C}^{*}$, is equal to
\begin{equation*}
f_{C^c \mid C}(\underline{\theta}_{C^c} \mid \underline{\theta}_{C}= \\ \underline{\theta}_{C}^{*}) =
\end{equation*}
\begin{equation}
\frac{\underline{c}^{(C^c \bigcup C)H}\left[
\left(\mathbb{I}_{\prod_{k \in C^c}(M_k+1)} \bigotimes_{k \in C}\underline{e}_{k}^{*}\right)
\left(\bigotimes_{k \in C^c} (\underline{e}_k\underline{e}_{k}^{H}) \right)
\left(\mathbb{I}_{\prod_{k \in C^c}(M_k+1)} \bigotimes_{k \in C}\underline{e}_{k}^{*H}\right) \right]\underline{c}^{(C^c \bigcup C)}}{f_{C^c}(\underline{\theta}_{C^c})},
\end{equation}
and the vector of parameters of the conditional distribution, $\underline{c}^{C^c \mid C}$, satisfies
\begin{equation}
\underline{c}^{(C^c \mid C)}=\frac{\left(\mathbb{I}_{\prod_{k \in C^c}(M_k+1)} \bigotimes_{k \in C}\underline{e}_{k}^{*H}\right)\underline{c}^{(C^c \bigcup C)}}{\sqrt{f_{C^c}(\underline{\theta}_{C^c})}}.
\end{equation}

\section{Characteristic Function and a Goodness-of-Fit Test}

Considering the results in Fan (1997) we applied a goodness-of-fit test based on the distance between the fitted characteristic function of the MNNTS model, which takes non-zero values on a finite set of integer vectors $\underline{t}$, and the empirical characteristic function makes it possible to construct a goodness-of-fit test for MNNTS models.
The characteristic function of a $d$-dimensional MNNTS model with parameter vectors $\underline{M}$ and $\underline{c}$ for the circular random vector $\underline{\Theta}$ is defined as
\[
\psi(\underline{t})=E(e^{i\underline{t}^\top\underline{\Theta}})
\]
which takes values on the vectors $\underline{t}$ with integer values such that, componentwise, $-\underline{M} \le \underline{t} \le \underline{M}$ with a total of $\prod_{k=1}^d (2M_k-1)$ different $\underline{t}$ vectors. For example, for a bivariate MNNTS model with $\underline{M}=(4,3)$ the characteristic function takes values on 63 different integer vectors $\underline{t}$ from $(-4,-3)$ to $(4,3)$. This is a consequence of the definition of the MNNTS density function. The characteristic function of a $d$-dimensional circular MNNTS vector $\underline{\Theta}$ for integer vectors $\underline{t}$ such that componentwise $-\underline{M} \le \underline{t} \le \underline{M}$ is given by
\[
\psi(\underline{t})=E(e^{i\underline{t}^\top\underline{\Theta}})=(2\pi)^d\sum \cdots \sum c_{k_1k_2 \cdots k_d}\bar{c}_{m_1m_2 \cdots m_d}
\]
where the $d$ sums are over indexes such that componentwise $\underline{k}-\underline{m}=-\underline{t}$ with $\underline{k}=(k_1,k_2,\ldots,k_d)$ and $\underline{m}=(m_1,m_2,\ldots,m_d)$. As expected, the characteristic function of the MNNTS model is a function of the elements of the $\underline{c}$ vector of parameters. The empirical characteristic function for a random sample of size $n$, $\underline{\theta}_1, \underline{\theta}_2, \ldots, \underline{\theta}_n$, is defined as
\[
C_n(\underline{t}^*)=\frac{1}{n}\sum_{j=1}^n e^{i\underline{t}^{*\top} \underline{\theta}_j}=\frac{1}{n}\sum_{j=1}^n \cos(\underline{t}^{*\top} \underline{\theta}_j)+
i\left(\frac{1}{n}\sum_{j=1}^n \sin(\underline{t}^{*\top} \underline{\theta}_j) \right).
\]
The goodness-of-fit test of Fan (1997) is based on the distance between the null and empirical characteristic functions in a set of evaluation points $\underline{t}^*$, $\underline{t}_1^*, \underline{t}_2^*, \ldots, \underline{t}_m^*$, where $m$ is the total number of evaluation points that is a function of the sample size $n$, $m=m_n$, that must satisfy $m \rightarrow \infty$ and $\frac{m^3}{n} \rightarrow 0$ as $n \rightarrow \infty$ to obtain a consistent test. Fan (1997) showed that in the case of a composite null hypothesis in which the parameters of the null model need to be estimated if the estimators are consistent, then the asymptotic distribution of the test statistic is normal. The test statistic $\hat{T}_n$ is a quadratic form on the vector of the average of the differences between the empirical and null characteristic functions for the different values of the $\underline{t}^*$ evaluation points. The test statistic $\hat{T}_n$ can be standardized to have a standard normal distribution by defining
\begin{equation}
\hat{Z}_n=\frac{n\hat{T}_n-2m}{2\sqrt{m}}
\end{equation}
where $m$ is the number of different evaluation points $\underline{t}^*$, $\underline{t}_1^*, \underline{t}_2^*, \ldots, \underline{t}_m^*$. In the case of the MNNTS model, as its characteristic function takes non-zero values in a finite set of discrete $\underline{t}$ vectors, then the goodness-of-fit test of Fan (1997) can be directly applied. As suggested by Fan (1997), with the identity matrix as the weight matrix, the goodness-of-fit test can be implemented by running a linear regression with a constant dependent variable equal to one and explanatory variables equal to the difference between the empirical characteristic function and the fitted null empirical characteristic function defined by the MNNTS model. The regression model is defined as $1=\underline{X}_j\underline{\gamma} + \epsilon_j$ for $j=1,2, \ldots, n$ where $\underline{\gamma}$ is the vector of unknown regression coefficients of dimensions $2m \times 1$ and $\underline{X}_j$ is the $j$-th row of the design matrix defined by
$\underline{X}_j=(\cos(\underline{t}_1^{*\top}\underline{\theta}_j)-Re(\psi(\underline{t}_1^*)),\ldots,\cos(\underline{t}_m^{*\top}\underline{\theta}_j)-Re(\psi(\underline{t}_m^*)),
\sin(\underline{t}_1^{*\top}\underline{\theta}_j)-Im(\psi(\underline{t}_1^*)),\ldots,\sin(\underline{t}_m^{*\top}\underline{\theta}_j)-Im(\psi(\underline{t}_m^*)))$ 
with $Re(a)$ and $Im(a)$ being the real and imaginary parts of the complex number $a$ and, $\epsilon_j$ being the regression random error. One can verify the $F$-test of the regression or can calculate $\hat{W}_n=\frac{nR^2-2m}{2\sqrt{m}}$ that converges in distribution to a standard normal random variable. The goodness of fit is rejected for large values of the test statistic $\hat{T}_n$ ($\hat{W}_n$). In our experience from simulations and given the constraint in the $\underline{c}$ parameter vector for an MNNTS model with vector $\underline{M} \ge 2$, the goodness of fit test should first be applied to $\underline{M}^*=\underline{M}-\underline{1}$. In a particular application of the goodness-of-fit test to the MNNTS model, we might have many explanatory variables with respect to the number of observations. In these situations, it is not possible to test for all the different non-zero values of the characteristic function corresponding to the discrete $\underline{t}$ vectors in accordance to the $\underline{M}$ vector, and a subset of these values should be used as evaluations points $\underline{t}^*$ in the test. In this sense, the goodness-of-fit test of the model is confirmed in a generally large subset of all the possible values of the support set of the null MNNTS characteristic function.

\section{Maximum Likelihood Estimation of MNNTS and an Alternative Estimation Method}

For $\theta_1, \theta_2, \ldots, \theta_n$, a random sample of univariate circular observations from an NNTS model with $M$ terms, the likelihood function is defined as follows:
\begin{equation}
L(\theta_1, \theta_2, \ldots, \theta_n \mid \underline{c})= \prod_{k=1}^n f(\theta_k \mid \underline{c}) = \prod_{k=1}^n \underline{c}^H \underline{e}\underline{e}^H \underline{c},
\label{likelihoodeq1}
\end{equation}
where $\underline{c}=(c_0, c_1, \ldots, c_{M})^\top$ and $\underline{e}_k=(1,e^{i\theta_k},e^{2i\theta_k}, \ldots, e^{Mi\theta_k})^\top$ for $k=1, 2, \ldots, n$.
Given the definition of an MNNTS density in terms of the Kronecker products in Equation \ref{mnntskronecker} and the definition of $\underline{\theta}=(\theta_1,\theta_2, \ldots, \theta_n)^\top$, the likelihood function in Equation \ref{likelihoodeq1} can be written as
\begin{equation}
L(\theta_1, \theta_2, \ldots, \theta_n \mid \underline{c}) = f_n(\underline{\theta} \mid \underline{c}^*) = \underline{c}^{*H} \left( \bigotimes_{k=1}^n \underline{e}_k\underline{e}_k^H \right) \underline{c}^{*} = \bigotimes_{k=1}^n \underline{c}^H \left( \bigotimes_{k=1}^n \underline{e}_k\underline{e}_k^H \right) \bigotimes_{k=1}^n \underline{c}
\label{likelihoodeq2}
\end{equation}
because of the independence of the observations in the random sample, $\underline{c}^*=\bigotimes_{k=1}^n \underline{c}$, with the normalizing constraint,
$\underline{c}^{*H}\underline{c}^{*} = \left( \frac{1}{2\pi} \right)^{n}$. If the likelihood function is maximized subject to the normalizing constraint,
the maximum likelihood estimator of parameter vector $\underline{c}^*$, $\hat{\underline{c}}^{*}_{ML}$, is an eigenvector of matrix $\bigotimes_{k=1}^n \underline{e}_k\underline{e}_k^H$ that satisfies
\begin{equation}
\hat{\underline{c}}^{*}_{ML} \propto \bigotimes_{k=1}^n \underline{e}_k,
\label{likelihoodeq3}
\end{equation}
implying $\bigotimes_{k=1}^n \hat{\underline{c}}_{ML} \propto \bigotimes_{k=1}^n \underline{e}_k$. Thus, the maximum likelihood is proportional to the Kronecker product of the $n$ trigonometric moments vectors. This result can be easily extended to a random sample of circular random vectors, $\underline{\theta}_1,\underline{\theta}_2, \ldots, \underline{\theta}_n$. Fern\'andez-Dur\'an and Gregorio-Dom\'inguez (2010) developed a numerical algorithm to obtain the maximum likelihood estimators in the univariate and
multivariate cases. When fitting the MNNTS model to a dataset, the univariate histograms of the components of the circular vector provide information about the number of modes in each component of the MNNTS circular distribution, which then allows us to determine the value of $\underline{M}$. The value of $M_k$ in the vector $\underline{M}$, corresponds to the maximum number of modes of the marginal distribution for the $k$-th component that can be confirmed by fitting a univariate NNTS model to the $k$-th element of the circular vector with different values of $M_k$ and identifying the best model in terms of the Bayesian Information Criterion (BIC). This rationale is used in the practical example in this paper. Alternatively, based on Equation \ref{likelihoodeq3}, a new estimator can be proposed by considering the minimization of the sum of the squared
distances of the estimator to the vectors of trigonometric moments, that is,
\begin{equation}
\hat{\underline{c}}_{MD} \propto \min_{\underline{c}^{**}} \sum_{k=1}^{n} || \underline{c}^{**} - \underline{e}_k ||^2.
\end{equation}
The solution for the estimator is based on minimizing the sum of squared distances, $\hat{\underline{c}}_{MD}$:
\begin{equation}
\hat{\underline{c}}_{MD} \propto \frac{1}{n}\sum_{k=1}^{n} \underline{e}_k.
\end{equation}
Thus, estimator $\hat{\underline{c}}_{MD}$ is proportional to the mean resultant of the vectors of the trigonometric moments. This result can be easily extended to the
multivariate case. In numerous simulation experiments, we confirmed that for very large sample sizes, $n$, $\hat{\underline{c}}_{MD}$, is a good approximation to $\hat{\underline{c}}_{ML}$. Clearly, $\hat{\underline{c}}_{MD}$ is significantly easier to obtain than $\hat{\underline{c}}_{ML}$ because it only involves calculating the mean resultant of the observed vectors of trigonometric moments and its subsequent normalization.

\section{Example}

\begin{figure}[h]
\begin{center}
\includegraphics[scale=1, bb= 0 0 504 504]{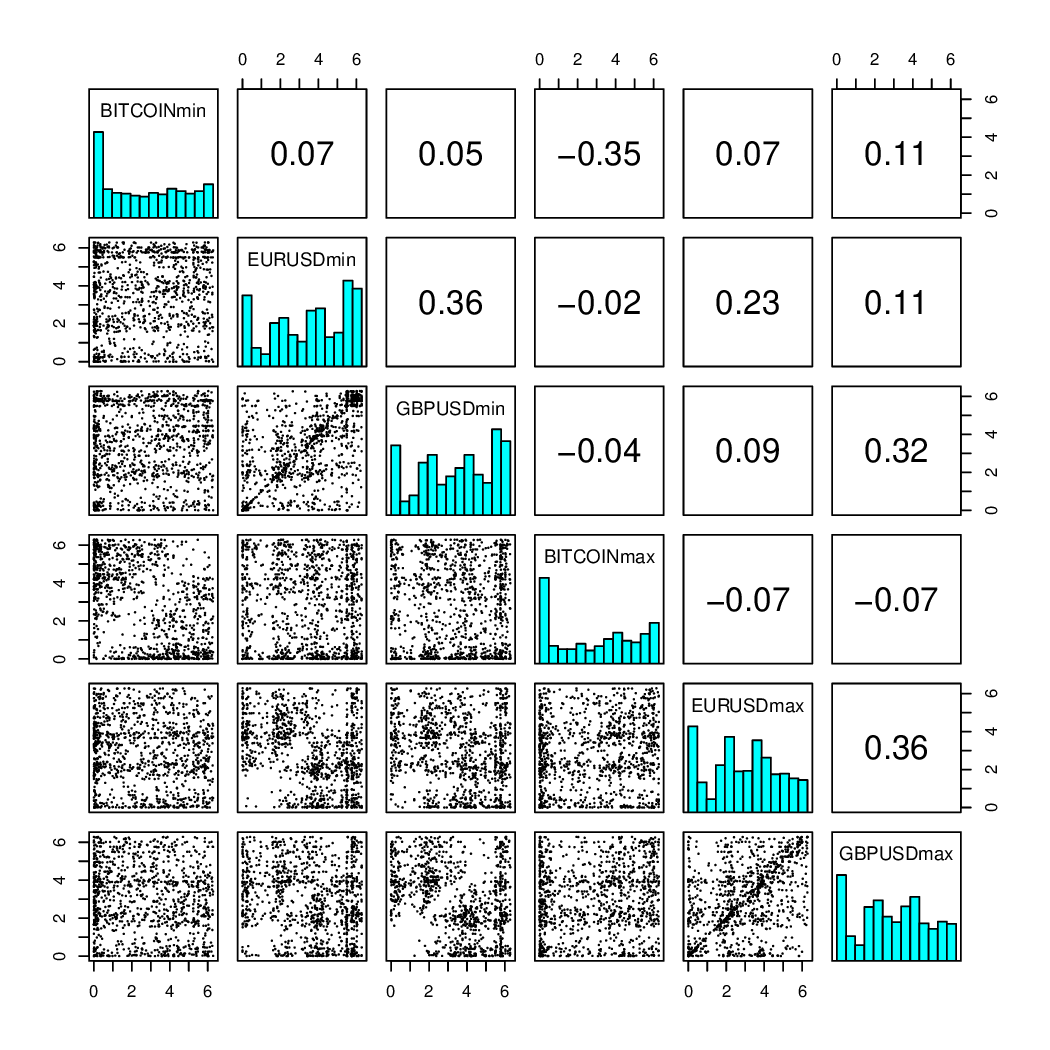}
\caption{Daily times of occurrence of maxima and minima of prices in financial markets: Lower diagonal half shows the bivariate dispersion plots. The histograms of each daily time of occurrence are shown in this half. The upper diagonal half shows the values of the circular correlation.}
\label{pairsandhistogramsminmax}
\end{center}
\end{figure}

\begin{figure}[h]
\begin{center}
\includegraphics[scale=1, bb= 0 0 504 504]{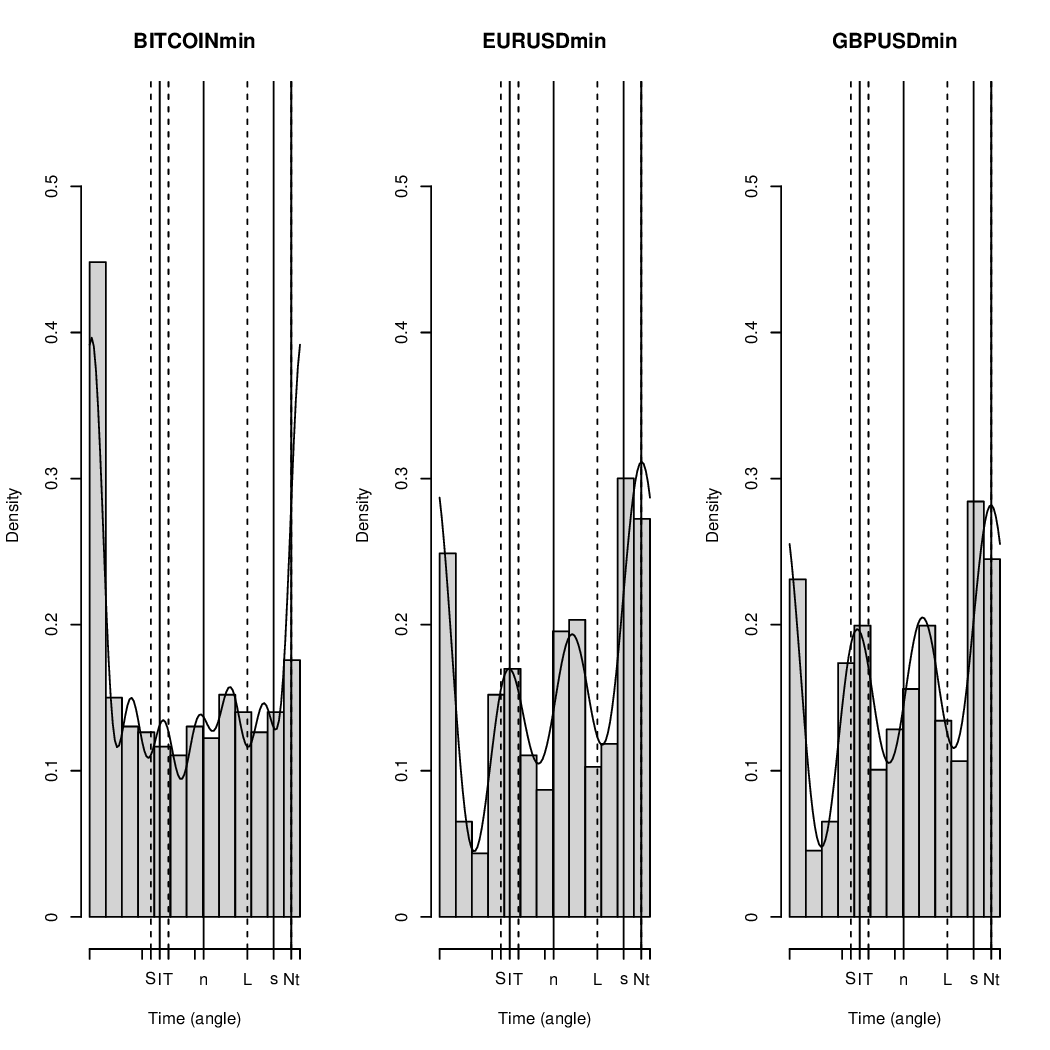}
\caption{Marginal densities of the daily times of occurrence of the minimum (BITCOINmin, EURUSDmin, and GBPUSDmin) that correspond to the mixtures of univariate NNTS densities. The solid vertical lines are the opening times of London (l), New York (n), Sidney (s) and Tokyo (t). The dashed vertical lines are the closing times of London (L), New York (N), Sidney (S) and Tokyo (T). The closing time of New York (N) and the opening time of Tokyo (t) are the same.}
\label{graphunivariatesmin}
\end{center}
\end{figure}

\begin{figure}[h]
\begin{center}
\includegraphics[scale=1, bb= 0 0 504 504]{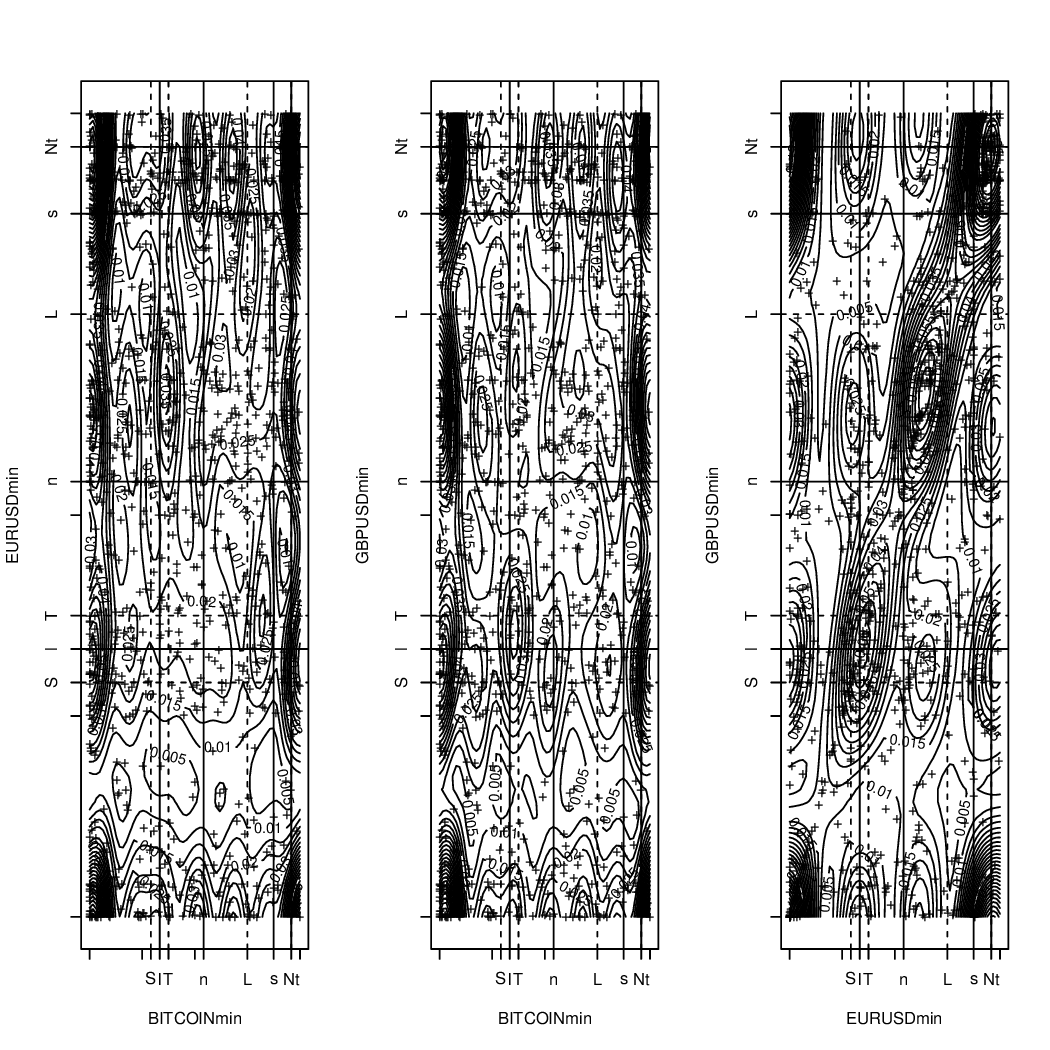}
\caption{Bivariate dispersion plots with the contours of the fitted bivariate densities of the daily times of occurrence of the minimum (BITCOINmin, EURUSDmin, and GBPUSDmin) derived from the fitted trivariate MNNTS model.
}
\label{graphbivariatesmin}
\end{center}
\end{figure}

We  examine the daily times of the maximum ask and minimum bid prices for Bitcoin, as well as the EURUSD and GBPUSD exchange rates. We downloaded Dukascopy's publicly available tick-by-tick data (Ntakaris 2018) from March 21, 2019, to March 21, 2023, yielding 1048 observations. The tick by tick data of Dukascopy covers bid and ask prices reported in the forex markets in London, New York, Sidney and Tokyo. The times are reported in milliseconds in GMT format and with approximately 20,000 events per day. Using the GMT format from April to October, the Tokyo forex market is open from 11pm to 9am GMT, Sidney from 9pm to 7am GMT, London from 8am to 6pm GMT and New York from 1pm to 11pm GMT. From November to March, the Tokyo forex market is open from 11pm to 9am GMT, Sidney from 10pm to 8am GMT, London from 7am to 5pm GMT and New York from 12pm to 10pm GMT. Once the datasets were downloaded, from the timestamp of the daily times of maximum ask prices and minimum bid prices in the datasets, the elapsed time in milliseconds since the beginning of the day was calculated and divided by the total number of milliseconds in a day to obtain the elapsed fraction of the day. The elapsed fraction of the day is then multiplied by 2$\pi$ in order to obtain the observed angle to be analyzed.
We refer to these angles as BITCOINmin, BITCOINmax, EURUSDmin, EURUSDmax, GBPUSDmin and GBPUSDmax. The estimates of the $\underline{c}$ vector of the parameters were obtained by applying the modified Newton algorithm on manifolds of Fern\'andez-Dur\'an and Gregorio-Dom\'inguez (2010 and 2016) and corresponds to the maximum likelihood estimates. Figure \ref{pairsandhistogramsminmax} shows the univariate histograms of the daily maxima and minima times of occurrence and bivariate dispersion plots. The circular correlation coefficients (Agostinelli and Lund 2017) are also included in Figure \ref{pairsandhistogramsminmax}. The pattern of the circular correlation coefficients shows that the BITCOIN (min and max) present the lowest correlations with the EURUSD and GBPUSD and that the highest correlations are among the EURUSD and GBPUSD. We are interested in fitting a trivariate model for the times of occurrence of the minimum bid prices (BITCOINmin, EURUSDmin, and GBPUSDmin) and a trivariate model for the times of occurrence of the maximum ask prices (BITCOINmax, EURUSDmax, and GBPUSDmax). By inspecting the univariate histograms and identifying the best BIC univariate NNTS model, we found that a suitable value of $M$ for the BITCOINmin and BITCOINmax is six, for the EURUSDmin and GBPUSDmin is three and, for the EURUSDmax and GBPUSDmax is four. Subsequently, a suitable model for the three minima is an MNNTS trivariate model with $\underline{M}=(6,3,3)$. Running the Fan (1997) goodness-of-fit test by using the auxiliary linear regression and considering the sample size of 1048 observations, we found that the fitted MNNTS model with $\underline{M}=(6,3,3)$ is validated for all the discrete $\underline{t}$ vectors such that $(-5,-2,-2) \le \underline{t} \le (5,2,2)$ for a total of 274 explanatory variables in the auxiliary regression. The F-test p-value of the regression was equal to 0.461 for the bid minima trivariate model. Figure \ref{graphunivariatesmin} presents the univariate histograms of the data with the corresponding univariate fitted densities derived from the fitted trivariate MNNTS model. The three fitted densities demonstrate an excellent fit to the univariate histograms. Note the correspondence between the modes of the fitted univariate densities and the opening and closing times of different markets, particularly London (from l to L), Tokyo (from t to T), and New York (from n to N). Figure \ref{graphbivariatesmin} includes the bivariate dispersion plots with the contours of the fitted bivariate densities derived from the fitted trivariate MNNTS model. Additionally, areas with high-density values occur around the opening and closing of the different financial markets. For instance, for the EURUSDmin and GBPUSDmin contour plot, the diagonal demonstrates a high concentration of the bivariate density function. The first row of Figure 4 displays the bivariate and marginal conditional densities of EURUSDmin and GBPUSDmin, given BITCOINmin values at the time London and Tokyo opened (New York close). This shows that the value of BITCOINmin has a slight effect on the joint and marginal densities of EURUSDmin and GBPUSDmin. This fact is reflected in the first eigenvalue (mixing probability) of the BITCOINmin being large and equal to 0.7710 in Table \ref{tableeigenvalues}, indicating a weak dependence of the EURUSDmin and GBPUSDmin on BITCOINmin. The second row of Figure 4 repeats the exercise, this time conditioning on the value of EURUSDmin to observe the effect on the conditional joint and marginal densities of BITCOINmin and GBPUSDmin. Here, the shape of the conditional joint and marginal densities of BITCOINmin and GBPUSDmin change considerably demonstrating a strong effect of EURUSDmin on BITCOINmin and GBPUSDmin. We repeat the trivariate analysis but for the maxima BITCOINmax, EURUSDmax, and GBPUSDmax, obtaining similar results to those of the minima. In contrast with the results for the minima, the fitted MNNTS model has $\underline{M}=(6,4,4)$, but as the minima model, it is validated at a 1\% significance level for all the discrete $\underline{t}$ vectors such that $(-5,-2,-2) \le \underline{t} \le (5,2,2)$ when applying the auxiliary regression of the Fan (1997) goodness-of-fit test with the F-test p-value equal to 0.022. Also, the BITCOINmax has a stronger dependence with EURUSDmax and GBPUSDmax as shown in Table \ref{tableeigenvalues} in which the first eigenvalue of BITCOINmax is equal to 0.5260. Finally, Figure \ref{graphresultantvector} includes the marginal densities estimated when applying the alternative mean resultant length estimator to the six angles simultaneously. For comparison purposes, the univariate NNTS maximum likelihood estimates are also included. The alternative mean resultant length estimator presents a reasonable fit to the univariate histograms. Future research will be conducted to determine the convergence properties of the alternative mean resultant length estimator that is computationally easy to implement and that can be used as an initial point in optimization algorithms to obtain the more complex maximum likelihood estimators. This example demonstrates the flexibility of MNNTS models when applied to real situations. The computations were done by using the R package CircNNTSRmult (Fern\'andez-Dur\'an and Gregorio-Dom\'inguez 2023).

\section{Conclusion}

Defining probability densities for multivariate circular data is a
challenging task. The MNNTS model developed by Fern\'andez-Dur\'an and Gregorio-Dom\'inguez (2014) is highly flexible, and the derivation of marginal and conditional
densities from the joint multivariate density is important when applying this
model, for instance, in time-series and spatial and spatiotemporal datasets involving
circular random variables. In this study, the essential algorithms for obtaining the marginal and conditional densities of any number of components of the joint
vector were specified. Furthermore, a goodness-of-fit test based on the empirical characteristic function can be applied to the MNNTS model, as its large number of parameters can be a disadvantage. In terms of its numerous parameters, the MNNTS model is comparable to computationally-demanding models such as neural networks.
The algorithms demonstrated substantial performance when applied to multidimensional real circular data.

\section*{Acknowledgements}

The authors wish to thank the Asociaci\'on Mexicana de Cultura, A.C. for its support.

\newpage
\renewcommand{\baselinestretch}{1.00}

\renewcommand{\baselinestretch}{1.00}
\begin{table}[t]
\begin{center}
\scalebox{.8}{
\begin{tabular}{|c|c|c||c|c|c|}
\hline
\multicolumn{6}{|c|}{Mixing Probabilities} \\
\hline
  BITCOINmin  & EURUSDmin  & GBPUSDmin   & BITCOINmax & EURUSDmax & GBPUSDmax  \\
\hline
0.7710   & 0.5691  & 0.5499  & 0.5260  & 0.5388  & 0.5517    \\
0.0982   & 0.2250  & 0.2083  & 0.2384  & 0.2195  & 0.2341    \\
0.0550   & 0.1470  & 0.1728  & 0.0969  & 0.1174  & 0.0929    \\
0.0297   & 0.0589  & 0.0690  & 0.0543  & 0.0764  & 0.0719    \\
0.0217   &         &         & 0.0443  & 0.0479  & 0.0494    \\
0.0171   &         &         & 0.0217  &         &           \\
0.0074   &         &         & 0.0184  &         &           \\
\hline
\end{tabular}}
\renewcommand{\baselinestretch}{1}
\caption{Mixing probabilities defining the marginal densities (see Equation \ref{marginaleq1}) of the daily times of occurrence of the minimum (first three columns) and the maximum (last three columns). The mixing probabilities correspond to the eigenvalues of the matrix in Equation \ref{eqmarginal2b}.
\label{tableeigenvalues}}
\end{center}
\end{table}

\begin{figure}[h]
\begin{center}
\includegraphics[scale=1, bb= 0 0 504 504]{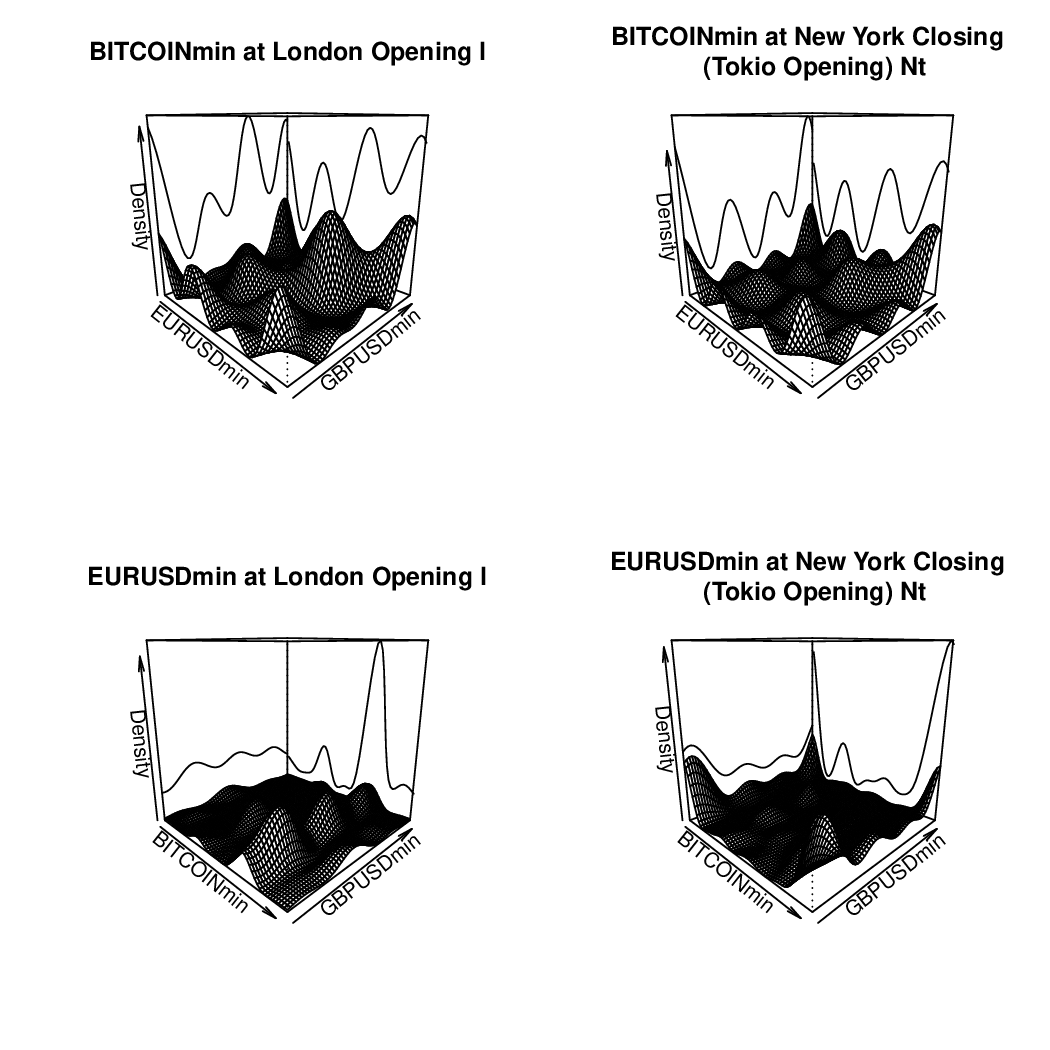}
\caption{First row plots: Bivariate and marginal conditional densities of EURUSDmin and GBPUSDmin, given BITCOINmin values at the time London l (left) and Tokyo t (right) opened (New York N close).
Last row plots: Bivariate and marginal conditional densities of BITCOINmin and GBPUSDmin, given EURUSDmin values at the time London l (left) and Tokyo t (right) opened (New York N close).}
\label{graphconditionalsmin}
\end{center}
\end{figure}

\begin{figure}[h]
\begin{center}
\includegraphics[scale=1, bb= 0 0 504 504]{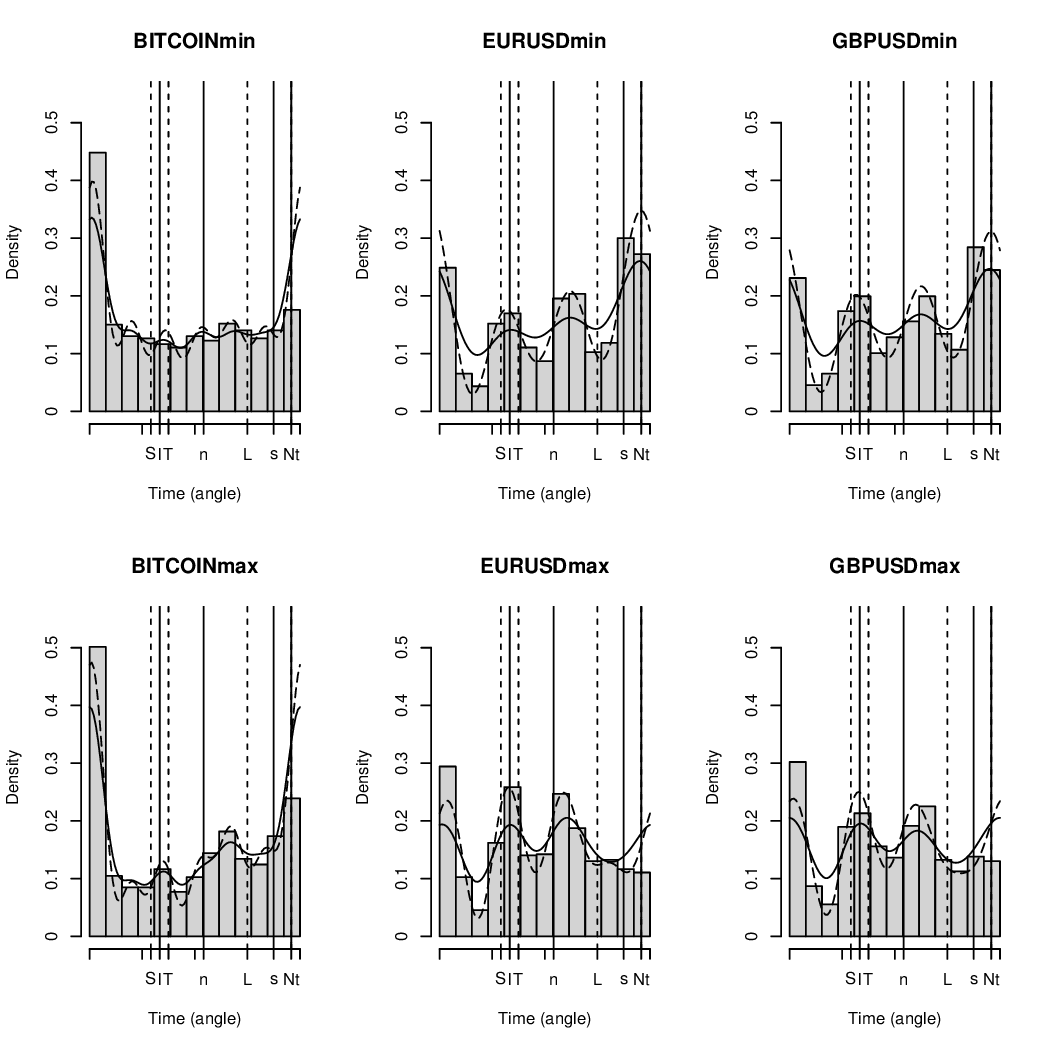}
\caption{Fitted marginal densities derived from the approximate estimator based on the resultant mean vector (solid line) when applying the alternative estimator to the six angles simultaneously. For comparison, the maximum likelihood (long dashed line) fitted univariate NNTS densities are also included.}
\label{graphresultantvector}
\end{center}
\end{figure}

\end{document}